\documentclass[%
 aip,
 jmp,%
 amsmath,amssymb,
preprint,%
]{revtex4-1}

\usepackage{graphicx}% Include figure files
\usepackage{dcolumn}% Align table columns on decimal point
\usepackage{bm}% bold math
\usepackage{graphicx}
\usepackage{natbib}
\usepackage{amsmath}
\usepackage{amssymb}
\usepackage{textcomp}
\usepackage{caption}
\usepackage{units}
\usepackage{subcaption}
\usepackage{epstopdf}

\begin{document}

%\preprint{AIP/123-QED}

\title{Dynamics and stability of a fluid filled cylinder rolling on an inclined plane}% Force line breaks with \\
\author{Rohit B. Supekar}
\affiliation{ Department of Mechanical Engineering, Indian Institute of Technology Madras, Chennai 600036, India}

\author{Mahesh V. Panchagnula}%
\email{mvp@iitm.ac.in}
\affiliation{Department of Applied Mechanics, Indian Institute of Technology Madras, Chennai 600036, India}

\date{\today}% It is always \today, today,

\begin{abstract}
	The dynamics and stability of a fluid-filled hollow cylindrical shell rolling on an inclined plane are analyzed. We study the motion in two dimensions by analyzing the interaction between the fluid and the cylindrical shell. An analytical solution is presented to describe the unsteady fluid velocity field as well as the cylindrical shell motion. From this solution, we show that the terminal state is associated with a constant acceleration. We also show that this state is independent of the liquid viscosity and only depends on the ratio of the shell mass to the fluid mass. We then analyze the stability of this unsteady flow field by employing a quasi-steady frozen-time framework. The stability of the instantaneous flow field is studied and transition from a stable to an unstable state is characterized by the noting the time when the eigenvalue crosses the imaginary axis. It is observed that the flow becomes unstable due to long wavelength axial waves. We find a critical Reynolds number ($\approx 5.6$) based on the shell angular velocity at neutral stability with the corresponding Taylor number being $\approx 125.4$.  Remarkably, we find that this critical value is independent of the dimensionless groups governing the problem. We show that this value of the critical Reynolds number can be explained from a comparison of time scales of motion and momentum diffusion, which predicts a value near $2 \pi$.
\end{abstract}

%Use showkeys class option if keyword
\maketitle

\section{Introduction}
Fluid motion induced by rotating walls is relevant to several applications. The problem of steady flow induced by rotating cylindrical walls has been addressed well in literature. \cite{Bentwich63, Ribando88, Jackson95, Meunier07} 
% analyzed the fluid flow in a semi-infinite rotating cylinder. \citet{Ribando88} studied the flow field in a partially filled tapered cylinder using laser-Doppler velocimetry technique. developed insights through experiments in the dynamics of a partially filled rolling cylinder. More recently, \citet{Meunier07} obtained experimental and theoretical results on the flow inside a weakly precessing and rotating cylinder. 
 These studies were all performed on a system where the cylindrical wall rotates at a constant angular velocity. In the current study, we analyze the dynamics of a system where the cylindrical wall accelerates. We consider the case of a hollow cylindrical shell fully filled with a viscous fluid and rolling down an inclined plane. The dynamical behavior of the cylindrical shell depends on the nature of the rotational velocity field and vice versa. In addition, the viscous dissipation as well as the terminal motion characteristics would both depend strongly on the fluid flow field. By solving the governing equations analytically, we determine velocity field in the fluid and examine the dependence of these two properties on the dimensionless parameters in the system. 

An important and interesting aspect of rotating flows is the multiple steady states and the stability characteristics of those states. Multiple states in steady rotating flows, especially those induced by rotating walls, have been studied by several researchers.\cite{Greenspan1969} The stability of these states has been classically a subject of interest. \cite{krueger1964,chen1967,ritchie1968,hall1975} Such studies generally lead to a critical Taylor number at which the flow becomes unstable. Fluid dynamics of rotating flows where the walls show time-varying motion have attracted less attention. Even in instances where the walls show time-varying motion, such a motion has been defined a priori. In contrast, we study a system where the motion of the wall and fluid are two-way coupled. One such example is the snail ball motion. \cite{balmforth2007} However, the stability of the fluid dynamic states in such two-way coupled systems is interesting and has not been studied. Two issues lead us to believe that the stability characteristics of such systems would be different from the case of steady rotating flows. Firstly, the history associated with the time varying wall motion has an effect on the base state as well as its stability. Secondly, the two-way coupled nature of the fluid dynamics suggests that a transition in the fluid dynamic state has an effect on the dynamical motion of the rotating wall. To the best of our knowledge, this is the first instance where the stability of an unsteady flow field in a two-way coupled system has been studied.

As mentioned before, we analyze the dynamics and stability of a rolling fluid-filled cylinder as a two-way coupled problem. We identify a critical Taylor number at which the flow becomes unstable. The angular velocity of the rotating wall is not enforced as an external input condition, but is developed as a result of its interaction of the cylinder with the fluid inside. Since the rotating wall accelerates, the fluid flow field is unsteady and we investigate its temporal stability using the frozen-time framework.\cite{MacKerrell2002} In this framework, the base flow is assumed to change much slower than the perturbation growth and hence the flow time is treated as a parameter in the system. We validate this approximation by presenting a comparison of growth rates of the Kinetic Energies in the base flow and the perturbations and showing that the base flow energy changes much slower than the rate of the growth of the energy in the perturbation mode.\cite{Yih68}
%The stability of the flow has been analyzed in the dimensionless parameter space and we find that the critical Taylor number is independent of these parameters. This is shown to be a result from a comparison of time scales associated with the motion and the viscous action, which practically is independent on how strong the gravitational field is or steep the inclined plane is. 

The manuscript is organized as follows. We begin with a mathematical description of the governing equations and present a generalized analytical solution in section \ref{GovEq}. Then, we apply this solution to the inclined plane problem in section \ref{IncPl}. We investigate the stability of this flow field in section \ref{LinStab} and finally discuss the implications of this work in section \ref{Conclusion}.

\section{Governing equations}
\label{GovEq}
Consider a rigid hollow cylindrical shell of mass, $M$, fully filled with an incompressible liquid, that is  imparted a rotation with  angular velocity, $\Omega(\tau)$. Here, $\tau$ denotes time. At $\tau=0$, the fluid inside the cylinder is at rest. The fluid flow field is governed by the Navier-Stokes equations in the cylindrical co-ordinate system $(r,\theta,z)$ in the usual notation ($z$ axis is placed along the axis of the cylinder). The velocity components in the respective directions are given by $(u,v,w)$. We assume that $w=0$ which results in $v=v(r)$ from the continuity equation. For this axisymmetric flow situation, the momentum equation reduces to,
\begin{subequations}
	\begin{equation}
	 \frac{v^2}{r} = \frac{\partial p}{\partial r}
	\label{PressureEqn}
	\end{equation}
	\begin{equation}
	\rho \frac{\partial v}{\partial \tau} = \mu \frac{\partial}{\partial r}\left(\frac{1}{r} \frac{\partial}{\partial r}\left(rv\right)\right)
	\label{PartialDiff}
	\end{equation}
	\label{orignialEqn}
\end{subequations}
Here, $\rho$ is the density of the fluid and $\mu$ is the dynamic viscosity of the fluid. The initial and boundary conditions are given by
\begin{equation}
\left. \begin{array}{ll}
v(r,0)=0 \\ 
v(0,t)=0 \\
v(R,\tau)=R\Omega(\tau)
\end{array} \right\}
\label{ICandBC}
\end{equation}

For the case of $\Omega(\tau)=\Omega_0$ (constant), the solution for the flow field has been given by \citet{batchelor67} as
\begin{equation}
v(r,\tau) = \Omega_0 r + 2\Omega_0 R \sum_{n=1}^{\infty}\frac{J_1\left(\lambda_n\frac{r}{R}\right)}{\lambda_n J_0(\lambda_n)} \exp{\left(-\lambda_n^2 \frac{\nu \tau}{R^2}\right)}
\label{VelocityField}
\end{equation}

Here, $R$ is the radius of the hollow cylinder and $\nu$, the fluid kinematic viscosity. $J_0(x)$ and $J_1(x)$ are the zeroth and first order Bessel's functions of the first kind. $\lambda_n$ is the $n^{th}$ zero of the equation, $J_1(x)=0$. 

\subsection{Time varying $\Omega$}
If $\Omega=\Omega(\tau)$, the solution to equation \eqref{PartialDiff} subject to the initial and boundary conditions given by equations \eqref{ICandBC} requires the use of Duhamel's theorem. For completeness, we state the Duhamel's theorem as given by \citet{Nellis09}:

``If  $T_f(\tau)$ is the response of a linear system to a single, constant non-homogeneous boundary condition of magnitude unity, then the response of the same system to a single, time-varying non-homogeneous boundary condition with magnitude $B(\tau)$ can be obtained from the following expression:
\begin{equation}
T(x,\tau) = \int_{\bar{\tau}=0}^\tau T_f(x,\tau-\bar{\tau})\frac{dB(\bar{\tau})}{d\bar{\tau}}d\bar{\tau} + B_0 T_f(x,\tau)
\label{DuhamelStatement}
\end{equation}	
Here, $B_0$ is the value of the function  $B(\tau)$ at $\tau=0$." This response to the unity magnitude boundary condition is called the fundamental solution. We now replace the single, time varying non-homogeneous boundary condition $B(\tau)$ with $R\Omega(\tau)$. In order to apply the theorem, we first need to identify the fundamental solution by letting $v(R,\tau)=R\Omega_0=1$ (with the constant $1$ having appropriate dimensions). Substituting this boundary condition into equation \eqref{VelocityField}, we obtain the following expression for the fundamental solution of the velocity field.
\begin{equation}
v_f(r,\tau) = \frac{r}{R} + 2 \sum_{n=1}^{\infty}\frac{J_1\left(\lambda_n\frac{r}{R}\right)}{\lambda_n J_0(\lambda_n)} \exp{\left(-\lambda_n^2 \frac{\nu \tau}{R^2}\right)}
\label{DuhamelMod}
\end{equation}

\noindent Now, setting $B(\tau) = v(R,\tau) = R\Omega(\tau)$ and realizing $B_0 = R\Omega_0$ ($\Omega_0$ is the initial angular velocity), we find from equations \eqref{DuhamelMod} and \eqref{DuhamelStatement} that,
\begin{multline}
v(r,\tau) = 
\int_{\bar{\tau}=0}^{\tau}\left\lbrace r+2R\sum_{n=1}^{\infty}\frac{J_1\left(\lambda_n\frac{r}{R}\right)}{\lambda_n J_0(\lambda_n)}
\exp{\left(-\lambda_n^2 \frac{\nu (\tau-\bar{\tau})}{R^2}\right)}
\right\rbrace \frac{d\Omega}{d\bar{\tau}}d\bar{\tau}
\\+\Omega_0\left\lbrace  r+2R\sum_{n=1}^{\infty}\frac{J_1\left(\lambda_n\frac{r}{R}\right)}{\lambda_n J_0(\lambda_n)} \exp{\left(-\lambda_n^2 \frac{\nu \tau}{R^2}\right)}\right\rbrace 
\label{VelocityEquation}
\end{multline}
Equation \eqref{VelocityEquation} describes the velocity field in the fluid, if the cylindrical shell is rotated with a time-varying angular velocity given by $\Omega(\tau)$. The first term on the $RHS$ of this equation is a contribution of the history associated with the wall motion, through the convolution integral. The corresponding pressure field can be obtained from equation \eqref{PressureEqn}. It may be noted that setting $\Omega(\tau)=\Omega_0$ in equation \eqref{VelocityEquation} allows us to recover equation \eqref{VelocityField}. The fluid inside the cylindrical shell exerts a torque on the shell and hence affects its dynamics. This will be the subject of the following discussion.

\subsection{Shear stress and torque}
We will denote the shear stress at the fluid-wall boundary by $\zeta$. We will use $T$ for the total torque exerted by the shear force at the wall. In this case, the instantaneous $\zeta$ would be given by:
\begin{equation} 
\zeta = \mu \left.\left\lbrace \frac{\partial v}{\partial r}-\frac{v}{r} \right\rbrace \right|_{r=R}
\label{ShearStress}
\end{equation}
\noindent On evaluating equation \eqref{ShearStress} using \eqref{VelocityEquation} and by invoking standard Bessel function identities, we obtain an expression for $\zeta$ and subsequently, an expression for $T$ defined as:
\begin{equation}
T(\tau)=\zeta(\tau) (2\pi Rh)R  =4\mu\pi Rh\  g(\tau)
\label{TorqueEquation}
\end{equation}
\noindent Here, $h$ is the length of the cylindrical shell. $g(\tau)$ in equation \eqref{TorqueEquation} is defined as
\begin{equation}
g(\tau) =  \sum_{n=1}^{\infty} \left\lbrace \int_{\bar{\tau}=0}^{\tau} \exp{(-k_n(\tau-\bar{\tau}))} \frac{d\Omega}{d\bar{\tau}}d\bar{\tau} +\Omega_0 \exp{(-k_n\tau)} \right\rbrace
\label{g(t)Expression}
\end{equation}
\noindent with $k_n$ given by
\begin{equation}
k_n = \lambda_n^2 \frac{\mu}{\rho R^2}
\label{KnEquation}
\end{equation}

\noindent It is to be noted that equation \eqref{TorqueEquation} is an expression for the torque exerted by the fluid on the hollow cylindrical shell in the most general case of $\Omega(\tau)$. We now seek the solution to the special case of the fluid-filled cylindrical shell rolling down an inclined plane.

\section{Cylinder rolling on an inclined plane}
\label{IncPl}

\begin{figure}
	\centering
	\includegraphics[height=60mm]{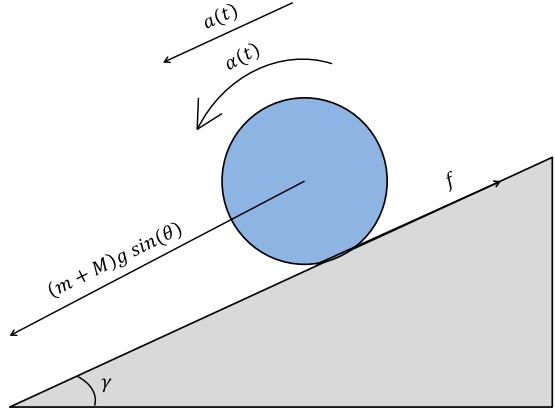}
	\caption{Schematic of the system showing the fluid-filled cylindrical shell}
	\label{cylinder_with_incline}
\end{figure}

We now consider the problem wherein a fluid-filled cylindrical shell is initially left at rest $(\Omega_0=0)$ on an inclined plane of inclination angle $\gamma$ in a uniform gravitational acceleration field denoted by $g$. We use $f$ to denote the force due to Coulomb friction between the cylinder and the inclined plane. We also assume that no slip occurs at this point of contact. See figure \ref{cylinder_with_incline} for a system schematic.

We write the mathematical equations describing the fluid motion in a (non-inertial) frame of reference moving with the center of mass of the cylinder as follows: 
\begin{subequations}
\begin{equation}
- \frac{v^2}{r} = \dot{\Omega}R\cos{(\theta-\gamma)} - g\sin{(\theta)} - \frac{\partial p}{\partial r}
\label{rmomentumincline}
\end{equation}
\begin{equation}
\frac{\partial v}{\partial \tau} = -\dot{\Omega}R\sin{(\theta-\gamma)} - g\cos{(\theta)} - \frac{1}{r}\frac{\partial p}{\partial \theta} + \nu \frac{\partial}{\partial r}\left(\frac{1}{r} \frac{\partial}{\partial r}\left(rv\right)\right)
\label{thetamomemtumincline}
\end{equation}
\label{NSinclinedplane}
\end{subequations}
Here, $v$ is the azimuthal velocity with respect to the center of mass of the cylinder. We write $p(r,\theta,\tau)$ as
\begin{equation}
p(r,\theta,\tau) = p_s(r,\theta,\tau) + p_d(r,\tau)
\label{PsandPd}
\end{equation}
where, 
\begin{equation}
p_s(r,\theta,\tau) = (\dot{\Omega}R\cos{(\theta-\gamma)} - g\sin{(\theta)})r 
\label{Ps}
\end{equation}
Substituting equations \eqref{PsandPd} and \eqref{Ps} in equations \eqref{NSinclinedplane}, we recover the original equations \eqref{orignialEqn} with $p$ replaced by $p_d$. We conclude that the fluid velocity field described by equation \eqref{VelocityEquation} remains valid, even in the accelerating frame of reference fixed to the center of mass of the moving cylinder.  
%Since the fluid is assumed to be incompressible, the pressure developed due to the pseudo force is balanced by the normal forces exerted by the shell walls. 

We write the equations describing the motion of a shell of mass $M$ with fluid of mass $m$ filling it completely. These equations are written in a stationary frame of reference. The balance of linear and angular momentum as well as the no slip condition can be stated mathematically as, 
\begin{equation}
\begin{array}{c}
(m+M)a=(m+M)g \sin{\gamma}-f\\
MR^2\frac{d\Omega}{d\tau}=fR-T\\
a= R\frac{d\Omega}{d\tau}
\end{array}
\label{dynamics1}
\end{equation}

\noindent Here, $a$ is the acceleration of the shell (and the fluid) center of mass. Eliminating $a$ and $f$ from equations \eqref{dynamics1}, we obtain,
\begin{equation}
(m+2M)\frac{d\Omega}{d\tau} = -T + (m+M)\bar{g}R
\label{diffequation}
\end{equation}

\noindent Here, $g\sin{\gamma}$ is replaced by $\bar{g}$ for brevity. Equation \eqref{diffequation} is rewritten by making use of equations \eqref{TorqueEquation} and \eqref{g(t)Expression} as follows:
\begin{equation}
\left( \frac{m+2M}{4\pi \mu h}\right) \frac{d\Omega}{d\tau}=-g(\tau)+\frac{(m+M)\bar{g}}
{4\pi \mu h} 
\label{finaleqn}
\end{equation}

\noindent It can be seen that equation \eqref{finaleqn} is a linear integro-differential equation, since $g(\tau)$ involves an integral of $\Omega(\tau)$. We now employ the method of Laplace Transforms to solve the equation. The motivation behind the usage of the method is that a convolution integral of the form in equation \eqref{g(t)Expression} is the multiplication of the functions in the $s$-domain. Hence, $G(s)$, which is the Laplace transform of $g(\tau)$ takes the following simplified form:
\begin{equation}
G(s) = \sum_{n=1}^{\infty} \frac{s W(s)}{k_n +s}
\label{G(s)}
\end{equation}

\noindent Here, $W(s)$ is the Laplace transform of $\Omega(\tau)$. Using equation \eqref{G(s)} and standard Laplace Transform properties, equation \eqref{finaleqn} can be transformed from time domain to $s$-domain as follows:
\begin{equation}
W(s) = \frac{(1+\pi_m)\left(\nicefrac{\bar{g}}{R}\right)}
{s^2\left\lbrace (1+2\pi_m)+4\sum_{n=1}^{\infty}\frac{1}{\lambda_n^2+s\widehat{\tau}}\right\rbrace}
\label{W(s)InclinedPlane}
\end{equation}

\noindent Here, $\pi_m$ is a dimensionless number given by
\begin{equation}
\pi_m = \frac{M}{\rho\pi R^2h}=\frac{M}{m}
\label{PIm}
\end{equation}
and $\widehat{\tau}$ is the viscous time scale defined as 
\begin{equation}
\widehat{\tau} = \frac{R^2}{\nu}
\end{equation}

\noindent We now attempt to find the analytical expression for $\Omega(\tau)$ from $W(s)$. This requires finding the inverse transform of $W(s)$. It is to be noted that $W(s)$ in equation \eqref{W(s)InclinedPlane} involves a series sum in the denominator. In order to evaluate the inverse transform, we truncate this series to a finite number of terms (say $N$) and denote the resulting Laplace function as $W_N(s)$. The inverse transform of $W_N(s)$ is then found analytically using Wolfram Mathematica\textsuperscript{\textregistered}.  The corresponding inverse transform, $\Omega_N(\tau)$ is of the form,
\begin{equation}
\Omega_N(\tau) = c_0 - \sum_{n=1}^{N}c_n \exp{(-a_n \tau)} + \alpha_N \tau
\label{Omega(t)}
\end{equation}

\noindent $c_n$ and $a_n$ are positive for all $n$. $\alpha_N$ is the estimate of the terminal angular acceleration (for finite $N$). As $n\to\infty$, $c_n\to 0$ and $a_n\to\infty$. Therefore, the series in equation \eqref{Omega(t)} is absolutely convergent. In addition, it was found that $N=15$ was sufficient through numerical convergence tests. In addition, we find  the terminal angular acceleration of the cylinder by using the Laplace Final Value Theorem. \cite{Ogata97} This leads to
\begin{equation}
\alpha_t = \left(\frac{2+2\pi_m}{3+4\pi_m}\right)\frac{\bar{g}}{R}
\label{alphaterminal}
\end{equation}

\noindent Finally, it was established that as $N\to\infty$, $\alpha_N = \nicefrac{d\Omega_N}{d\tau}$ computed from equation \eqref{Omega(t)} approaches $\alpha_t$ in equation \eqref{alphaterminal}. 

\begin{figure}
	\centering
	\includegraphics[height=60mm]{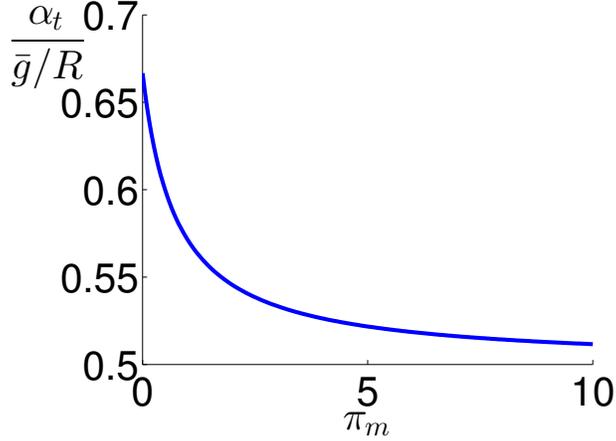}
	\caption{Non-dimensional terminal angular acceleration versus $\pi_m$. $\nicefrac{\alpha_t}{\bar{g}R}$ tends to the following two limits:  $\nicefrac{2}{3}$ (similar to a rigid cylinder), when $\pi_m\rightarrow 0$ and $\nicefrac{1}{2}$ (similar to a hollow rigid shell), when $\pi_m\rightarrow \infty$}
	\label{AlphaVsPIm}
\end{figure}
\begin{figure}
	\centering
	\includegraphics[height=60mm]{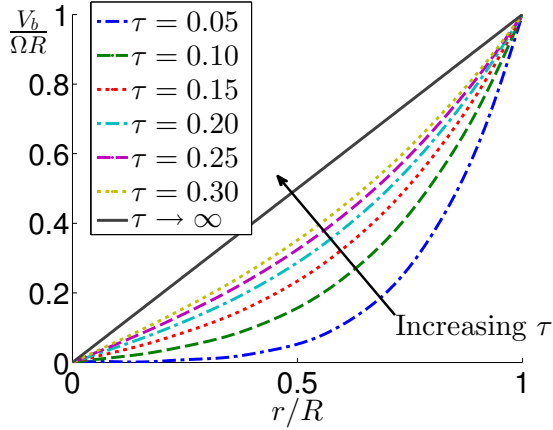}
	\caption{The azimuthal velocity of the fluid ($V_b$, non-dimensionalised with $\Omega(\tau)$) versus the non-dimensional radius}
	\label{vel_profiles}
\end{figure}

It is interesting that the terminal acceleration of the fluid-filled cylindrical shell given by equation \eqref{alphaterminal} lies between the two limits corresponding to a solid cylinder and a hollow cylinder respectively. These limits are given by $\nicefrac{\bar{g}}{2R}$ and $\nicefrac{2\bar{g}}{3R}$ are the accelerations of a rigid cylinder and a hollow cylinder, respectively. Therefore, the terminal state of a fluid-filled shell is similar to a solid cylinder when $\pi_m \to 0$ and to a hollow cylinder when $\pi_m \to \infty$. In addition, it is interesting to note that the terminal acceleration is independent of viscosity. This can be explained physically from the fact that viscosity plays a dual role in the system. Firstly, it is responsible for the diffusion of momentum towards the centerline, which takes the velocity field reach the asymptotic non-dimensional profile as shown in figure \ref{vel_profiles}. In this role, it is responsible for a reduction in the torque the fluid applies on the cylindrical shell. In the second role, it is responsible for the inhibiting torque itself, since the shear stress at the wall is directly proportional to the viscosity. These two effects appear to counteract, resulting in a terminal state that is viscosity independent. However, the unsteady dynamics for finite $\tau$, does depend on the fluid viscosity. 
\begin{figure}
	\centering
	\includegraphics[height=60mm]{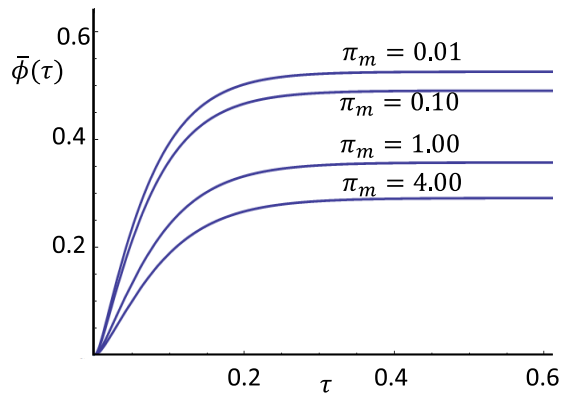}
	\caption{The total dissipation rate versus time for varying values of $\pi_m$. As $\pi_m$ increases, the terminal dissipation rate decreases, indicating a tendency of the fluid to achieve a velocity profile closer to rigid body rotation, since the mass of the fluid is lower.}
	\label{dissipation_vs_time}
\end{figure}

The difference between rigid body motion and the current problem lies in the fact that the fluid inside the shell is a source of energy dissipation. Dissipation is dependent on the shear stress, which is directly proportional to the shear rate given by $\nicefrac{\partial v}{\partial r}-\nicefrac{v}{r}$. The dissipation rate can be estimated from the series solution obtained in equation \eqref{VelocityEquation}, specialized for the case of a cylinder rolling on an inclined plane. The total dimensionless rate of dissipation as a function of dimensionless time is given by 
\begin{equation}
\bar{\phi}(\tau) = \frac{4}{(\nu /R)^2} \int_{0}^{R}r\left\lbrace \frac{\partial v}{\partial  r} - \frac{v}{r}\right\rbrace^2 dr
\label{DissipationRate}
\end{equation}

Figure \ref{dissipation_vs_time} shows the plot of dimensionless dissipation rate versus dimensionless time for different values of $\pi_m$. Here, $\bar{g}R^3/\nu^2=5$, which is a non-dimensional quantity defined in the next section. We note that the asymptotic dissipation rate is higher for lower values of $\pi_m$. This also implies that the dissipation rate is higher for higher values of terminal acceleration. It is to be noted that the asymptotic dissipation rate is non-zero but interestingly, plays no role in determining the terminal acceleration given by equation \eqref{alphaterminal}. These observations of the dissipation characteristics also have a bearing on the stability of the flow field, which is considered next.

\section{Linear stability analysis}
\label{LinStab}
We now consider the second objective of this study i.e., the stability of the flow field generated inside a fluid-filled cylindrical shell as it rolls down an inclined plane. We note that the base velocity field computed in the previous section is two-dimensional and unsteady. We employ the ``frozen-time'' approximation \cite{MacKerrell2002}, an approach where time is treated as a parameter that determines the instantaneous flow field. The stability of this base flow is studied by introducing perturbations and studying their growth using linear stability theory. Our aim is to determine the time at which the velocity profile in the fluid becomes unstable. Finally, we would like to study the nature of the system at neutral stability, in the space of the dimensionless parameters. We define the base velocity field given in equation \eqref{VelocityEquation} for the case of a cylinder rolling on an inclined plane as follows: 
\begin{equation}
V_{b_*}(r_*,\tau_*) = \mathcal{L}^{-1}\left\lbrace r_*W_*(s_*)+2\sum_{n=1}^{N} \frac{J_1(\lambda_n r_*)}{\lambda_n J_0(\lambda_n)} \frac{s_* W_*(s_*)}{\lambda_n^2 +s_*}   \right\rbrace
\label{BaseVelField}
\end{equation}

Here, subscript $_*$ denote non-dimensional quantities, which have been non-dimensionalized using the following scales:	
\begin{subequations}
	\begin{equation}
	\left. \begin{array}{ll}
	r_*=r/R \\
	\tau _*=  \tau/\widehat{\tau} \\ 
	s_* = \widehat{\tau}s \\
	\Omega_* = \widehat{\tau}\Omega
	\end{array} \right\}
	\end{equation}
	\label{NonDimensionalQuantities}
	\begin{equation}
	V_* = \frac{V}{\nicefrac{\nu}{R}},\  \text{(Here, $V_*$ is any velocity.)}
	\label{V*}
	\end{equation}
\end{subequations}
\noindent The dimensionless parameters that govern the system dynamics are
\begin{subequations}
	\begin{equation}
	\pi_m = \frac{M}{m} = \frac{M}{\rho\pi R^2 h}
	\label{PIm}
	\end{equation}
	\begin{equation}
	\pi_g = \frac{g\sin{\gamma}R^3}{\nu^2} = \frac{\bar{g}R^3}{\nu^2}
	\label{PIg}
	\end{equation}
\end{subequations}

\noindent Henceforth, we will drop subscript $_*$ and treat all the variables as dimensionless, unless specified otherwise. Hence finally, $W(s)$ is obtained from equation \eqref{W(s)InclinedPlane} as
\begin{equation}
W(s) = \frac{(1+\pi_m)\pi_g}{s^2\left\lbrace 1+2\pi_m+4\sum_{n=1}^{N}\frac{1}{\lambda_n^2+s}\right\rbrace}
\end{equation}

\subsection{Perturbation equations} 
The velocity field given in equation \eqref{BaseVelField} is perturbed and its stability studied. We denote the perturbation quantities with a $\widetilde{(\ )}$ and the (unperturbed) base flow quantities with $b$ as the subscript. As before, we will use $\tau$ to denote the flow time (which is now reduced to a parameter) and $t$ to denote a fast time co-ordinate along which the perturbation is allowed to grow or decay. The perturbed flow variables can now be written as,
\begin{equation}
\left. \begin{array}{ll}
p(r,\theta,z,t) = p_b(r,\theta,z;\tau)+\widetilde{p}(r,\theta,z,t) \\
\vec{V}(r,\theta,z,t) = \vec{V_b}(r,\theta,z;\tau)+\vec{\widetilde{V}}(r,\theta,z,t) \\
\end{array} \right\}
\label{PerturbationQuantities}
\end{equation}

\noindent Note that here, $\vec{V_b} \equiv (0,V_b(r;\tau),0)$. Substituting equation \eqref{PerturbationQuantities} in the Navier-Stokes equations along with the continuity equation, and then eliminating the terms which satisfy the base flow after linearization leads to
\begin{subequations}
	\begin{equation}
	\nabla.\vec{\tilde{V}}=0
	\label{PerturbationContinuity}
	\end{equation}
	\begin{equation}
	\frac{\partial  \vec{\tilde{V}}}{\partial  t} + \vec{V_b}.\nabla \vec{\tilde{V}} + \vec{\tilde{V}}.\nabla \vec{V_b} = \frac{-\nabla \tilde{p}}{\rho} + \nu \nabla^2 (\vec{\tilde{V}})
	\label{PerturbationNavier}
	\end{equation}
\end{subequations}

\noindent It is to be noted that the time derivative in equation \eqref{PerturbationNavier} only involves $\vec{\tilde{V}}$ as $\nicefrac{\partial \vec{V_b}}{\partial t} \ll \nicefrac{\partial\vec{\tilde{V}}}{\partial t}$, owing to the frozen time approximation. It can be shown that the perturbation can be resolved into normal modes of the form,
\begin{equation}
(\tilde{p},\vec{\tilde{V}})=(\widehat{p}(r), \vec{\widehat{V}}(r))\exp{[\sigma t + i(\alpha z+\beta \theta)]}
\label{NormalMode}
\end{equation}

\noindent It is to be noted that $\vec{\widehat{V}}(r) = [\widehat{u}(r),\widehat{v}(r),\widehat{w}(r)]$, where $\widehat{u}$, $\widehat{v}$ and $\widehat{w}$ are the eigenfunctions corresponding to perturbation velocity in the $r$, $\theta$ and $z$ directions, respectively. Here, $(\alpha,\beta)$ are the wavenumbers in the $(z,\theta)$ directions. It is to be noted that $\alpha$ is a real number in our case, while $\beta$ is an integer. In addition, it is to be noted from equation \eqref{NormalMode} that the stability is studied by imposing three-dimensional disturbances on the two-dimensional flow field. Substituting equation \eqref{NormalMode} into equations \eqref{PerturbationContinuity} and \eqref{PerturbationNavier}, we obtain the following system of linear ordinary differential equations in dimensionless form. For convenience of notation, we have dropped the $\widehat{(\ )}$ symbol. It is also to be noted that henceforth, all the variables are treated to be dimensionless and only functions of $r$.
\begin{subequations}
	\begin{equation}
	\frac{du}{dr} + \frac{u}{r} + \frac{i \beta}{r}v + i\alpha w = 0
	\end{equation}
	\begin{equation}
	\frac{d}{dr} \left( \frac{1}{r}\frac{d}{dr}(ru)\right) +\left\lbrace -\left( \frac{\beta^2}{r^2}+\alpha^2\right)+\frac{2V_b}{r}-\frac{i \beta}{r}V_b\right\rbrace u 
	-\frac{2i \beta}{r^2}v -\frac{dp}{dr} = \sigma u
	\end{equation}
	\begin{equation}
	\frac{d}{dr} \left( \frac{1}{r}\frac{d}{dr}(rv)\right)+ \left\lbrace -\left( \frac{\beta^2}{r^2}+\alpha^2\right)-\frac{i \beta}{r}V_b\right\rbrace v 
	+\left(\frac{2i \beta}{r^2}-\frac{V_b}{r}-\frac{dV_b}{dr} \right)u -i \beta p = \sigma v 
	\end{equation}
	\begin{equation}
	\frac{1}{r}\left(\frac{d}{dr}\left(r\frac{dw}{dr}\right)\right) + \left\lbrace -\left( \frac{\beta^2}{r^2}+\alpha^2\right)-\frac{i \beta}{r}V_b\right\rbrace w 
	-i \alpha p = \sigma w
	\end{equation}
	\label{PertLin}
\end{subequations}

\noindent The corresponding boundary conditions are 
\begin{subequations}
	\begin{equation}
	u=v=w=0 \text{ at } r=1
	\label{noslipBC}
	\end{equation}
	\begin{equation}
	u=v=\frac{dw}{dr}=\frac{dp}{dr}=0 \text{ at } r=0
	\label{symmetryBC} 
	\end{equation}
	\label{BounCLin}
\end{subequations}
\noindent Here,  $\sigma$ has been non-dimensionalized by $1/\widehat{\tau}$.  Boundary conditions mentioned in equations \eqref{noslipBC} and \eqref{symmetryBC} are obtained from no slip and symmetry arguments, respectively. Note that, $V_b$ is the base velocity field (function of $r$ and $\tau$) and is already known from equation \eqref{BaseVelField}. The above equations \eqref{PertLin} and \eqref{BounCLin} constitute an eigenvalue problem with $\sigma$ being the eigenvalue governing the stability of the system. Typically, the system is deemed to be unstable if the real part of $\sigma$ is greater than $0$ for any value of $(\alpha,\beta)$. We will present more rigorous arguments for the case of time-varying velocity fields, later.

The eigenvalue problem described above was solved numerically. We have used finite difference method for this purpose. The $r$ co-ordinate ($0\le$$r$$\le$$1$) is discretized using $\mathcal{N}$ points.  For the first derivatives, we have chosen a forward differencing scheme and the second derivatives are approximated by a central differencing scheme. The eigenvalue problem was solved using MATLAB$^\circledR$ to obtain the eigenvalues.

A brief discussion of the numerical challenges is in order at this point. The resulting discretized equations are of the form $AV = \sigma B V$. Here $A$ and $B$ are sparse square matrices of size $4\mathcal{N}$ and $V$ is the column vector of size $4\mathcal{N}$. The number of eigenvalues obtained scales with $\mathcal{N}$. Therefore, one of the challenges lies in differentiating between the true eigenvalues of equations \eqref{PertLin} and the spurious ones which arise out of the spatial discretization. One way to eliminate the spurious ones is by checking the value's convergence with increasing $\mathcal{N}$. Since the eigenvalue with the largest real part determines the stability of the system, we apply this principle to that one alone and have ensured that it was non-spurious. Henceforth, the eigenvalue with the largest real part would be termed as the maximum eigenvalue ($\sigma$).

\begin{figure}
	\centering
	\includegraphics[height=60mm]{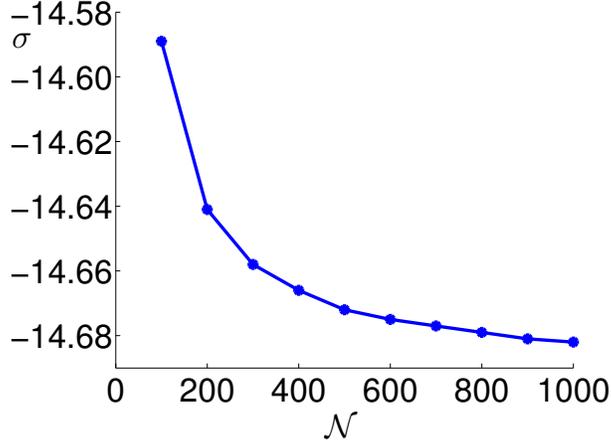}
	\caption{Largest eigenvalue($\sigma$) vs $\mathcal{N}$ for $\beta=0, \alpha=0.1$ and $\tau=6.5$. The eigenvalue is converged to within 0.1\% with about $300$ nodes indicating grid independence at this $\mathcal{N}$.}
	\label{converging_eig}
\end{figure}

It is desired that the maximum eigenvalue for a given $(\alpha,\beta)$ is independent of $\mathcal{N}$. In order to demonstrate this, we consider the largest eigenvalue for $(\alpha,\beta)=(0.1,0)$ for varying $\mathcal{N}$. Figure \ref{converging_eig} is a plot showing this variation. It is clear from this plot that a value of $\mathcal{N}=300$ is sufficient to ensure convergence of the largest eigenvalue to within $0.1\%$. Higher accuracy may be obtained by using greater values of $\mathcal{N}$ which comes at an increased computational cost. For the remainder of the work presented herein, $\mathcal{N}=300$ was chosen as being sufficient to generate grid-independent estimates of the maximum eigenvalues. Incidentally, the maximum eigenvalue (as well as the other non-spurious eigenvalues) of the system were real in all the cases investigated during this study.

\subsection{Neutral stability characteristics}
As mentioned before, we wish to find the time ($\tau^*$) at the onset of instability for the case where a fluid-filled cylindrical shell is released from rest. Equation \eqref{BaseVelField} describes the time-varying velocity field in the fluid. It may be recalled that time is being treated as a parameter in the velocity field description. The time instant at which the largest growth rate for all possible $(\alpha,\beta)$ changes sign from being negative to a positive value is typically noted as the point of neutral stability. We wish to discuss the underlying assumptions based on the arguments presented by \citet{Yih68}. The dimensionless kinetic energy in the base velocity field ($\mathcal{K}_b$) and the perturbation velocity field (over unit length of the cylinder) are respectively given by: 
\begin{subequations}
	\begin{equation} 
	\mathcal{K}_b = \iiint_{\mathbf{\mathcal{V}}}^{} \frac{{V_b}^2}{2} d \mathcal{V}
	\label{baseKE}
	\end{equation}
	\begin{equation}
	\widehat{\mathcal{K}} = \iiint_{\mathbf{\mathcal{V}}}^{} \frac{{\mid{\vec{\widetilde{V}}}}\mid ^2}{2} d \mathcal{V}
	\label{perturbKE}
	\end{equation}
\end{subequations}

\noindent In order to characterize the growth of kinetic energy in the base flow, we define an amplification factor, $\sigma_b$ where 
\begin{equation}
2 \sigma_b=\frac{1}{\mathcal{K}_b}\frac{d \mathcal{K}_b}{d\tau}
\label{ampsigb}
\end{equation}

\noindent It is to be noted  from the definition in equations \eqref{perturbKE} and \eqref{NormalMode} that
\begin{equation}
2\sigma =\frac{1}{\widehat{\mathcal{K}}}\frac{d \widehat{\mathcal{K}}}{dt}
\label{ampsigp}
\end{equation}

\begin{figure}
	\includegraphics[height=60mm]{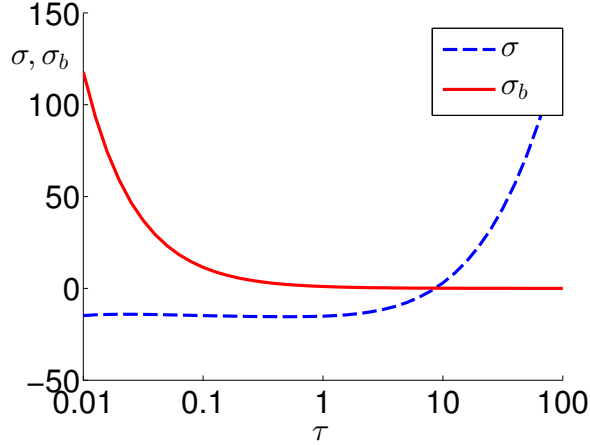}
	\caption{The variation of $\sigma_b$ and $\sigma$ with $\tau$ for $\pi_m=10^{-3}$ and $\pi_g=1$ ($\alpha=0.01$ and $\beta=0$). It can be noted that at $\tau^* \approx 8.1$, $\sigma=\sigma_b$, indicating the point of neutral stability. At this $\tau^*$, $\sigma=\sigma_b \approx 0.12$.}
	\label{sig_sigb}
\end{figure}

\noindent Incidentally, $\sigma_b$ denotes an exponential growth rate of the kinetic energy in the base flow analogous to $\sigma$ for the perturbation. When the amplification of kinetic energy in the perturbation mode is greater than the rate of growth of kinetic energy in the base flow ($\sigma>\sigma_b$), instability is said to have set in. At the point of neutral stability, $\sigma=\sigma_b$. This is a more accurate description of the neutrally stable point than to simply require that $\sigma=0$. We note the time instant, $\tau^*$, at which $\sigma=\sigma_b$. Figure \ref{sig_sigb} is a plot of $\sigma$ and $\sigma_b$ versus $\tau$ over four orders of magnitude of $\tau$. As can be seen from this figure, $\sigma_b$ is initially very high. However, at the cross-over point, $\sigma=\sigma_b\approx 0.12$ with $\tau^* \approx 8.1$. Further, since the value of $\sigma_b$ is small at this point, the use of the quasi-steady frozen flow approximation is validated. 
\begin{figure}
	\centering
	\includegraphics[height=60mm]{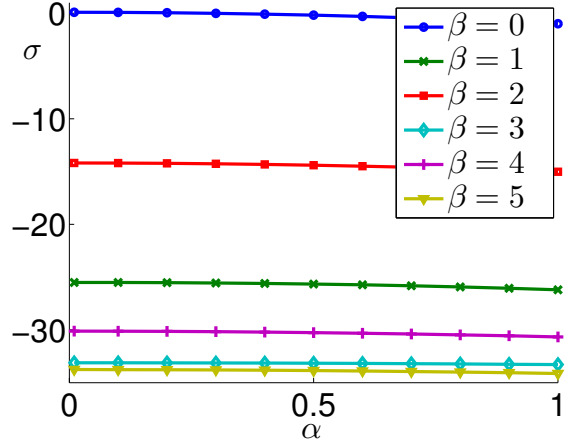}
	\caption{Maximum eigenvalue ($\sigma$) versus $\alpha$  for different values of $\beta$ for $\pi_m=0.001$, $\pi_g =1$ and $\tau =8.1$.}
	\label{eig_vs_alpha_and_beta}
\end{figure}
\begin{figure}
	\centering
	\includegraphics[height=60mm]{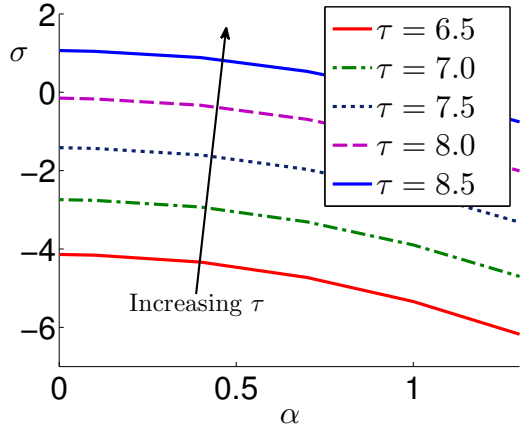}
	\caption{Maximum eigenvalue ($\sigma$) versus $\alpha$  for $\beta=0$ for $\pi_m=0.001$ and $\pi_g$ =1. The maximum eigenvalue changes sign between $\tau=8.0$ and $8.5$ indicating that the system is at neutrally stable conditon for a $\tau$ between these two values.}
	\label{eig_vs_alpha}
\end{figure}

For steady flows, $\sigma_b=\nicefrac{d \mathcal{K}_b}{d \tau}=0$. For unsteady (accelerating) flows, $\nicefrac{d \mathcal{K}_b}{d \tau}>0$.  We have assumed that the flow is quasi-steady and undergoing small values of accelerations (say, for a gently sloping inclined plane). It is to be noted from equation \eqref{Omega(t)} that the angular acceleration is an exponentially decreasing function of $\tau$. This further suggests that the frozen time approximation remains acceptable for the duration of the cylinder's motion.

As mentioned before, we would like to identify a pair $(\alpha,\beta)=(\alpha^*,\beta^*)$, where the real part of the growth rate, $\sigma$ is maximum. Towards this end, we find the value of $\sigma$ for different $(\alpha, \beta)$ pairs, at different values of $\tau$.

Figure \ref{eig_vs_alpha_and_beta} is a plot of the maximum eigenvalue versus $\alpha$ for different values of $\beta$ at $\tau=8.1$, which was identified as the neutrally stable point in time in figure \ref{sig_sigb}. This plot corresponds to the case where $\pi_m=0.001$, $\pi_g=1$. It is evident that the maximum eigenvalue occurs in the case of $\beta=0$ for all values of $\alpha$. As further validation, we consider a plot of $\sigma$ versus $\alpha$ for different values of $\tau$. Firstly, it is to be noted  that as $\tau$ is increased, the largest eigenvalue always corresponds to the case of $\beta=0$. The neutral stability transition happens at for $\alpha^* \rightarrow 0$. This appears to suggest that the system exhibits Type $III_s$ instability as classified by \citet{CrossHohen93}. This further implies that the system is susceptible to long wavelength spanwise waves, which have also been observed in other similar rotating flows.  

\begin{figure}
	\centering
	\includegraphics[height=60mm]{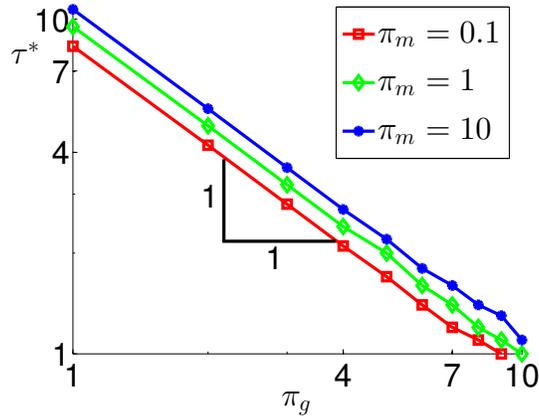}
	\caption{Plot of $\tau^*$ against $\pi_g$. It can be seen that $\tau^* \pi_g=f(\pi_m)$.}
	\label{log_log_pig}
\end{figure}

We now turn our attention to characterizing $\tau^*$ as a function of the dimensionless parameters, $\pi_m$ and $\pi_g$. Bisection algorithm was used to identify $\tau^*$ for each case, within an error of $0.1$. $\pi_g$ was varied between $1$ and $10$ and  $\pi_m$ was varied from $10^{-2}$ to $10^2$. The variation of $\tau^*$ versus $\pi_g$ is shown in figure \ref{log_log_pig}.

Two key observations can be drawn from figure \ref{log_log_pig}. Firstly, as $\pi_m$ is increased (at a constant value of $\pi_g$), the time to instability increases (albeit slightly). This implies that the higher inertia associated with the shell is a stabilizing factor. It can be noted from figure \ref{dissipation_vs_time} that a high value of $\pi_m$ corresponds to a lower rate of dissipation. Hence, the system which dissipates lesser is found to be more stable.  Secondly, $\tau^*$ is inversely proportional to $\pi_g$ implying that $\tau^* \pi_g$ is a constant. Therefore,
\begin{equation}
\tau^* \pi_g=f(\pi_m)
\label{prod_eq}
\end{equation} 
$\pi_m$ and $\pi_g$ are as stated in equations \eqref{PIm} and \eqref{PIg}. \noindent Defining,  $V_s=g'\tau_i$ ($\tau_i$ is the dimensional time at neutral stability) as the corresponding velocity scale, we find that the $LHS$ of equation \eqref{prod_eq} can be rewritten in the form of critical Reynolds number, $Re_c$ defined as
\begin{equation}
\tau^* \pi_g=\frac{V_s R}{\nu} = Re_{c}
\label{RecDef}
\end{equation}

\begin{figure}
	\centering
	\includegraphics[height=60mm]{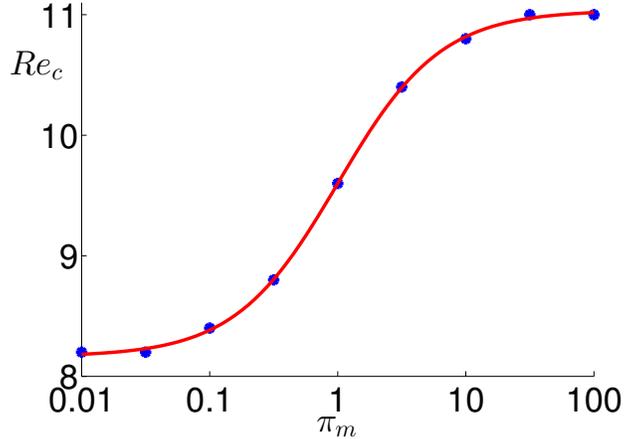}
	\caption{Critical Reynolds number against $\pi_m$. Symbols indicate exact computations from the linear stability analysis and the solid line indicates a least squares fit of a sigmoid function of the form $Re_c=A+B\tanh{(C \log{\pi_m})}$. Physically, the inflection point at $\pi_m=1$ and the limits of $Re_c$ as $\pi_m \rightarrow 0,\infty$ are to be noted.}
	\label{Rec_vs_log(pim)}
\end{figure}

From equation \eqref{prod_eq}, we find that $Re_c$ is only a function of $\pi_m$, which we will investigate next. Figure \ref{Rec_vs_log(pim)} is a plot of $Re_c$ versus $\pi_m$. Firstly, it is seen that $Re_c$ is a weak function of $\pi_m$; four orders of magnitude variation in $\pi_m$ causes $\pm 15\%$ variation in $Re_c$. A least squares fit to the data points in this figure suggests $9.60+1.44\tanh{(1.25\log{(\pi_m)})}$ as a good fit to the data, which has been plotted in figure \ref{Rec_vs_log(pim)} as the solid line. It is important to note the significance of this curve. For a given ratio of the masses of the cylindrical shell and the fluid contained as well as the inclination angle, the time of the onset of instability can be determined from this critical Reynolds number. 

It would be useful to study the dynamical characteristics of the cylinder at the onset of instability. Towards this end, we consider a Reynolds number, $Re_\Omega$, based on the angular velocity at the point of neutral stability, $\Omega^*$, where
\begin{equation}
Re_{\Omega} = \frac{(\Omega^* R)R}{\nu}
\end{equation}

$Re_\Omega$ defined above and $Re_c$ defined in equation \eqref{RecDef} would be equivalent if $\Omega$ was linearly dependent on $\tau$. Since this is not the case (in the most general instance), we have chosen to define an instantaneous actual Reynolds number (based on the wall motion) at the inception of instability. For the range of $\pi_m$ (four orders of magnitude) and $\pi_g$ (two orders of magnitude) studied herein, $Re_\Omega$ was found to be independent of both these dimensionless groups and equal to $5.6 \pm 0.1$. This is remarkable in that the onset of instability based on the instantaneous angular velocity appears to occur at a universal transition value. The corresponding critical value of the Taylor number (defined most commonly as $Ta=\frac{4\Omega^2 R^4}{\nu^2}$) is $125.4$. 

This value of $Re_\Omega$ can be reconciled using time scale arguments. The time scale associated with the shell motion, when rotating at an angular velocity, $\Omega$, is given by $\nicefrac{2\pi}{\Omega}$. The time scale associated with momentum diffusion from the shell is given by $\nicefrac{R^2}{\nu}$. As $\Omega$ increases from rest, a critical point is reached when these two time scales are comparable. At this critical point, $\nicefrac{2\pi}{\Omega^*}  \sim \nicefrac{R^2}{\nu}$. Rearranging terms, we find that $Re_\Omega = \frac{\Omega^* R^2}{\nu} \sim 2\pi$. This value of $Re_\Omega \approx 2\pi$ is remarkably close to the actual value of $5.6$ obtained from the exact calculations. Therefore, it can be concluded that the flow is stable as long as the time scale associated with the addition of momentum by the wall motion (to a thin layer of fluid near the wall) is more than the momentum diffusion time scale. It becomes unstable when this condition is breached. The smaller momentum diffusion time scale also ensures nearly solid body rotation in the fluid at shorter time instants, again implying stability.

\section{Conclusion}
\label{Conclusion}
The dynamics of a fluid-filled cylindrical shell rolling down an inclined plane has been discussed. It was shown that the two-dimensional fluid velocity field affects the unsteady dynamics of the rolling cylinder and vice versa. The velocity field as well as the cylinder motion were described by solving the governing integro-differential equations analytically, using the method of Laplace Transforms. From this analytical solution, it was shown that the system reaches a state of terminal angular acceleration at long times. The system dynamics are shown to be governed by two dimensionless groups: (i) $\pi_m$, which is the ratio of the masses of the liquid to the outer shell and (ii) $\pi_g$, which is the ratio of gravitational to viscous time scales respectively. The former group was found to determine the terminal characteristics, while the latter group controls the unsteady dynamics leading to the terminal state.

We then examined the stability of the velocity field in the fluid using linear instability theory by invoking the quasi-steady frozen time approximation. In this stability analysis, time is treated as a parameter, rendering the flow quasi-steady. The analytical solution describing the velocity field was perturbed in three-dimensions and the linearized equations were investigated to describe the perturbation growth. The resulting eigenvalue problem for the growth rate was solved numerically. From this analysis, it was found that the largest eigenvalue (maximum growth rate) becomes positive (first) for the case of axial perturbation (in preference to azimuthal perturbation). In addition, the time at which the perturbation growth rate exceeds the amplification factor of energy in the mean flow field, is termed the neutral stability time. This neutral stability time is studied as a function of $\pi_g$ and $\pi_m$. These curves indicated that an increased inertia on the side of the shell or increased viscosity of the fluid, both have a stabilizing effect on the system, by delaying the onset of instability. It was also found that a lower dissipative system was more stable, analogous to the fact that instability and pattern formation occur at high dissipation rates in nonequilibrium systems. Finally, a Reynolds number based on the instantaneous angular velocity at the neutral stability condition was defined. It was shown that this critical Reynolds number takes on a universal value, which is approximately equal to $5.6$ and is independent of both $\pi_m$ and $\pi_g$. This critical value is rationalized using scaling arguments based on the motion time scale and momentum diffusion time scale to be a value close to $2\pi$.

\begin{acknowledgments}
	RBS thanks Jason R. Picardo for a careful reading and useful suggestions for improving the manuscript.
\end{acknowledgments}

\bibliography{pof_manuscript}
%\nocite{*}

\end{document}